\newcommand{\roughly}[1]%
    {{\mathrel{\raise.3ex\hbox{$#1$\kern-.75em\lower1ex\hbox{$\sim$}}}}}

\documentclass[12pt]{article}
\usepackage{paper2e}
\usepackage{latexsym}
\usepackage{graphicx}

\begin{document}
\begin{titlepage}

\preprint{IPMU09-0041}

\title{Scale-invariant cosmological perturbations from
 Ho\v{r}ava-Lifshitz gravity without inflation}

\author{Shinji Mukohyama}

\address{
Institute for the Physics and Mathematics of the Universe (IPMU)\\ 
The University of Tokyo\\
5-1-5 Kashiwanoha, Kashiwa, Chiba 277-8582, Japan
}

\begin{abstract}
 Based on the renormalizable theory of gravitation recently proposed by 
 Ho\v{r}ava, we present a simple scenario to generate almost
 scale-invariant, super-horizon curvature perturbations. The anisotropic
 scaling with dynamical critical exponent $z=3$ implies that the
 amplitude of quantum fluctuations of a free scalar field generated in
 the early epoch of the expanding universe is insensitive to the Hubble
 expansion rate and, thus, scale-invariant. Those fluctuations are later
 converted to curvature perturbations by the curvaton mechanism or/and
 the modulated decay of heavy particles/oscillating fields. This
 scenario works, for example, for power law expansion $a\propto t^p$
 with $p>1/3$ and, thus, does not require inflation. Also, this scenario
 does not rely on any additional assumptions such as the detailed
 balance condition. 
\end{abstract}

\end{titlepage}

\section{Introduction}
\label{sec:introduction}

Ho\v{r}ava~\cite{Horava:2009uw} recently proposed a renormalizable
theory of gravitation at a Lifshitz point as a candidate for ultraviolet
(UV) completion of general relativity. The essence of the theory is the 
following anisotropic scaling with dynamical critical exponent $z=3$: 
%
\begin{equation}
 \vec{x}\to b \vec{x}, \quad t \to b^z t.
  \label{eqn:scaling}
\end{equation}
In the infrared (IR), due to relevant deformation by lower-dimensional
operators, the theory flows to $z=1$ and general relativity is
recovered. Cosmology in this theory was investigated in
refs.~\cite{Takahashi:2009wc,Calcagni:2009ar,Kiritsis:2009sh,Lu:2009em}.

The purpose of this paper is to present a scenario of generation of
scale-invariant cosmological perturbations, based on Ho\v{r}ava-Lifshitz
gravity. Before going into details in the next section, let us explain
the basic idea.

The $z=3$ scaling implies that a dispersion relation for a physical
degree of freedom in the UV should be of the form 
%
\begin{equation}
 \omega^2 \propto k_{phy}^6 = \frac{k^6}{a^6},
\end{equation}
where $k_{phy}$ is the physical momentum, $k$ is the comoving momentum 
and $a$ is the scale factor of the background Friedman-Robertson-Walker
(FRW) universe. When $\omega^2\gg H^2$, a mode with $k$ does not feel
the expansion of the universe and oscillates. On the other hand, when
$\omega^2\ll H^2$, the Hubble friction acts efficiently and the mode
freezes. Therefore, if 
%
\begin{equation}
 \partial_t\left(a^6H^2\right) > 0,
  \label{eqn:freeze-condition}
\end{equation}
then a mode with a given $k$ initially oscillates but freezes out
afterwards. In quantum theory, this implies generation of super-horizon
quantum fluctuations. The condition more precise than
(\ref{eqn:freeze-condition}) will be shown later but it agrees with
(\ref{eqn:freeze-condition}) for power law expansion 
$a\propto t^p$. Indeed, both conditions give $p>1/3$.

Under the scaling (\ref{eqn:scaling}) with general $z$, a scalar field
$\Phi$ should scale as 
%
\begin{equation}
 \Phi \to b^s \Phi, \quad s = -\frac{3-z}{2}. 
\end{equation}
Therefore, the amplitude of quantum fluctuations is expected to scale as 
%
\begin{equation}
 \delta\Phi \propto E^{-s/z}=E^{(3-z)/(2z)}, 
\end{equation}
where $E$ is the typical energy scale of the problem. In cosmology, $E$
is set by the Hubble expansion rate $H$ and thus we obtain 
%
\begin{equation}
 \delta\Phi \propto H^{(3-z)/(2z)}. 
\end{equation}
The standard result $\delta\Phi\propto H$ in relativistic theories is 
recovered by setting $z=1$. By setting $z=2$, we recover the result in
the ghost inflation 
$\delta\Phi\propto H^{1/4}$~\cite{ArkaniHamed:2003uy,ArkaniHamed:2003uz}. 
On the other hand, in the UV limit of Ho\v{r}ava-Lifshitz gravity,
i.e. at $z=3$, the amplitude of quantum fluctuation is insensitive to
the Hubble expansion rate and, thus, scale-invariant.

The rest of this paper is organized as follows. In Sec.~\ref{sec:scalar}
we consider a free scalar field in Ho\v{r}ava-Lifshitz gravity and show
that the amplitude of quantum fluctuations is insensitive to the Hubble
expansion rate and, thus, scale-invariant.  In Sec.~\ref{sec:curvaton}
we invoke the curvaton mechanism or/and the modulated decay of heavy 
particles/oscillating fields to convert quantum fluctuations of the
scalar field to curvature perturbations. Finally, Sec.~\ref{sec:summary}
is devoted to a summary of this paper and discussions.

\section{Free scalar field in Ho\v{r}ava-Lifshitz gravity}
\label{sec:scalar}

Let us consider a free scalar field $\Phi$ in Ho\v{r}ava-Lifshitz 
gravity. In the UV, the action for $\Phi$ should respect the scaling
(\ref{eqn:scaling}) with dynamical critical exponent $z=3$. Since the
scaling (\ref{eqn:scaling}) treats the time coordinate and the spatial 
coordinates unequally, it is convenient to adopt the ADM decomposition 
of the metric~\cite{Arnowitt:1962hi}, 
%
\begin{equation}
 ds^2 = -N^2 dt^2 + q_{ij}(dx^i+N^idt)(dx^j+N^jdt), 
\end{equation}
where $q_{ij}$ is the $3$-dimensional spatial metric, $N$ is the lapse
function, and $N^i$ is the shift vector. Assuming that the time kinetic
term is canonical, the action for $\Phi$ should be of the form 
%
\begin{equation}
 I = \frac{1}{2}\int dt d^3\vec{x} a^3 N\sqrt{q}
  \left[\frac{1}{N^2}\left(\partial_t\Phi-N^i\partial_i\Phi\right)^2 
   + \Phi {\cal O}\Phi \right],
  \label{eqn:free-scalar}
\end{equation}
where 
%
\begin{equation}
 {\cal O} = \frac{1}{M^4}\Delta^3 
  - \frac{\lambda}{M^2}\Delta^2
  + \Delta - m^2,
  \label{eqn:O-Delta}
\end{equation}
$\Delta$ is the Laplacian associated with the spatial metric $q_{ij}$,
$M$ and $m$ are mass scales and $\lambda$ is a dimensionless
constant. Here, the coefficient of $\Delta$ in (\ref{eqn:O-Delta}) is
set to unity so that the speed of propagation in the IR fixed point
agrees with the speed of light, which is unity in our unit. Note that
the coefficient of $\Delta^3$ in (\ref{eqn:O-Delta}) is required to be
positive by the $z=3$ scaling and stability in the UV.

For quantum fluctuations generated in a sufficiently early epoch of the
expanding universe, where the Hubble expansion rate $H$ is much greater
than $M$, $M/\sqrt{|\lambda|}$ and $m$, physical wavelengths of the
corresponding modes are so short that only the highest-order spatial
derivative term is important. Thus, in the very early universe, we can
set 
%
\begin{equation}
 {\cal O} = \frac{1}{M^4}\Delta^3.
\end{equation}
In the flat FRW background
%
\begin{equation}
 ds^2 = -dt^2 + a(t)^2\delta_{ij}dx^idx^j,
\end{equation}
the action in the UV limit is 
%
\begin{equation}
 I_{UV} = \frac{1}{2}\int dt d^3\vec{x} a^3
  \left[(\partial_t\Phi)^2
   + \frac{1}{M^4a^6}\Phi(\delta^{ij}\partial_i\partial_j)^3\Phi 
  \right].
\end{equation}
In this regime, the action depends only derivatively on
$\Phi$. Therefore, the value of $\Phi$ averaged over a comoving size
corresponding to the present horizon is generically non-zero:
%
\begin{equation}
 \langle\Phi\rangle\ne 0.\label{eqn:averagePhi}
\end{equation}
In the following we shall investigate quantum fluctuations $\delta\Phi$
around the average: $\Phi=\langle\Phi\rangle+\delta\Phi$. 

In order to investigate quantum fluctuations, it is convenient to use
the conformal time $\eta$ defined by $dt=a d\eta$ so that 
%
\begin{equation}
 I_{UV} = \frac{1}{2}\int d\eta d^3\vec{x} 
  \left[a^2 (\partial_{\eta}\delta\Phi)^2
   + \frac{1}{M^4a^2}\delta\Phi(\delta^{ij}\partial_i\partial_j)^3
   \delta\Phi  \right].
\end{equation}
Noting that the Klein-Gordon norm for this system is
%
\begin{equation}
 \left(\delta\Phi_1,\delta\Phi_2\right)_{KG} = 
  -i\int d\vec{x}^3 a^2
  \left( \delta\Phi_1 \partial_{\eta}\delta\Phi_2^*
   - \delta\Phi_2^*\partial_{\eta}\delta\Phi_1 \right),
\end{equation}
the normalized mode function is
%
\begin{equation}
 \phi_{\vec{k}} = \frac{e^{i\vec{k}\cdot\vec{x}}}
  {(2\pi)^3} \times 2^{-1/2}k^{-3/2}M
  \exp \left(-i\frac{k^3}{M^2}\int \frac{d\eta}{a^2}\right).
  \label{eqn:mode-function}
\end{equation}
Here, this mode function is chosen so that its short-time behavior is
the same as the positive-frequency mode function in Minkowski
background~\footnote{The corresponding vacuum state minimizes the
Hamiltonian of the system.} and that it has the norm
$(\phi_{\vec{k}},\phi_{\vec{k}'})_{KG} 
=\delta^3(\vec{k}-\vec{k}')/(2\pi)^3$. 
Note that $\phi_{\vec{k}}$ approaches a constant value in the
$a\to\infty$ limit, if the integral 
%
\begin{equation}
 \int^{\eta_{\infty}} \frac{d\eta}{a^2}
  =\int^{t_{\infty}} \frac{dt}{a^3}
  \label{eqn:integral}
\end{equation}
converges. Here, $\eta_{\infty}$ and $t_{\infty}$ are values of $\eta$
and $t$, respectively, in the limit $a\to\infty$. For example, the power  
low expansion 
%
\begin{equation}
 a \propto t^p, \quad p> 1/3 
\end{equation}
satisfies this condition. Under the condition that the integral
(\ref{eqn:integral}) converges, a mode function stops oscillating and
freezes out when $\omega=M^{-2}k^3/a^3$ becomes comparable to
the Hubble expansion rate $H$. Note that the corresponding wavelength at
the freeze-out is far longer than the Hubble horizon radius, $1/H$,
because of the unusual dispersion relation, $\omega^2=M^{-4}k_{phy}^6$,
in the UV ($H\gg M$). This is the reason why super-horizon fluctuations
can be generated without accelerated expansion of the universe.

By expanding the field $\delta\Phi$ as
%
\begin{equation}
 \delta\Phi = \int d^3\vec{k}
  \left(\phi_{\vec{k}} a_{\vec{k}}
   + \phi^*_{\vec{k}} a^{\dagger}_{\vec{k}}
  \right),
\end{equation}
the operators $a_{\vec{k}}$ and $a^{\dagger}_{\vec{k}}$ satisfy
%
\begin{equation}
 \left[a_{\vec{k}}\ , \ a^{\dagger}_{\vec{k}'} \right]
  = (2\pi)^3 \delta^3(\vec{k}-\vec{k}'), \quad
 \left[a_{\vec{k}}\ , \ a_{\vec{k}'} \right] = 
 \left[a^{\dagger}_{\vec{k}}\ , \ a^{\dagger}_{\vec{k}'} \right]
 = 0.
\end{equation}
The power spectrum ${\cal P}_{\delta\Phi}$ is defined by 
%
\begin{equation}
 \langle 0| \delta\Phi_{\vec{k}}\delta\Phi_{\vec{k}'}|0\rangle
  = (2\pi)^3\delta^3(\vec{k}+\vec{k}') P_{\delta\Phi},
\end{equation}
and 
%
\begin{equation}
 P_{\delta\Phi} = \frac{2\pi^2}{k^3}{\cal P}_{\delta\Phi},
\end{equation}
so that 
%
\begin{equation}
 \langle 0|\delta\Phi^2|0\rangle = \int \frac{dk}{k}{\cal P}_{\delta\Phi}. 
\end{equation}
Here, $k\equiv \sqrt{\vec{k}^2}$, 
%
\begin{equation}
 \delta\Phi_{\vec{k}} \equiv
  \int d\vec{x}^3 e^{i\vec{k}\cdot\vec{x}}\delta\Phi, 
\end{equation}
and the vacuum $|0\rangle$ is defined by 
%
\begin{equation}
 a_{\vec{k}}|0\rangle = 0,  \quad {}^\forall \vec{k}. 
\end{equation}
By using the mode function (\ref{eqn:mode-function}) we obtain
%
\begin{equation}
 {\cal P}_{\delta\Phi}^{1/2} = \sqrt{\frac{k^3}{2\pi^2}}
  \left|(2\pi)^3\phi_{\vec{k}}\right|
  = \frac{M}{2\pi}. 
\end{equation}
This is insensitive to the Hubble expansion rate and, thus,
scale-invariant.

So far, we considered the UV limit of Ho\v{r}ava-Lifshitz gravity and
showed the scale-invariance of quantum fluctuations in generic expanding
flat FRW background, provided that the integral (\ref{eqn:integral})
converges.

On the other hand, in the IR the theory flows to the $z=1$ fixed
point, where the local Lorentz invariance is recovered. Therefore, for
quantum fluctuations generated at sufficiently low Hubble expansion
rate, i.e. when $H\ll \min(M, M/\sqrt{|\lambda|}, m)$, the standard
result $\delta\Phi\propto H$ should hold. Therefore, for those
fluctuations generated at the IR epoch, the spectrum can be almost
scale-invariant only if the Hubble expansion rate is almost constant,
i.e. only if inflation takes place. Moreover, if fluctuations of today's
cosmological scales are to be generated at the IR epoch, then
accelerated expansion is necessary not only for the scale-invariance of
quantum fluctuations but also to stretch causally generated modes to
super-horizon sizes. The spectrum is in general red-tilted since the
Hubble expansion rate decreases as the universe expands.

In intermediate epoch, the amplitude of quantum fluctuations depends on
the Hubble expansion rate but the dependence is not as strong as in the
IR epoch. The higher the Hubble expansion rate is, the weaker the
dependence of the amplitude on the expansion rate is. Therefore, for
quantum fluctuations generated in sufficiently but not too much early
epoch of the expanding universe, the spectrum depends only modestly on
the background FRW evolution and, as a result, is only slightly tilted.

\section{Curvaton mechanism or/and modulated decay}
\label{sec:curvaton}

In the previous section we have shown that quantum fluctuations of a
free scalar field is almost scale-invariant for generic expanding flat
FRW backgrounds. However, in order to account for temperature
anisotropies observed in the cosmic microwave background and density
perturbations observed in the large scale structure of the universe,
fluctuations of the scalar field must be converted to curvature
perturbations. For this purpose we invoke the curvaton
mechanism~\cite{Lyth:2001nq,Enqvist:2001zp,Moroi:2001ct} or/and the 
modulated decay of heavy particles/oscillating fields. The latter is 
similar to the modulated reheating~\cite{Dvali:2003em,Kofman:2003nx}.

\begin{figure}
 \begin{center}
\includegraphics[angle=0,trim = 0 0 0 0 ,scale=0.5, clip]{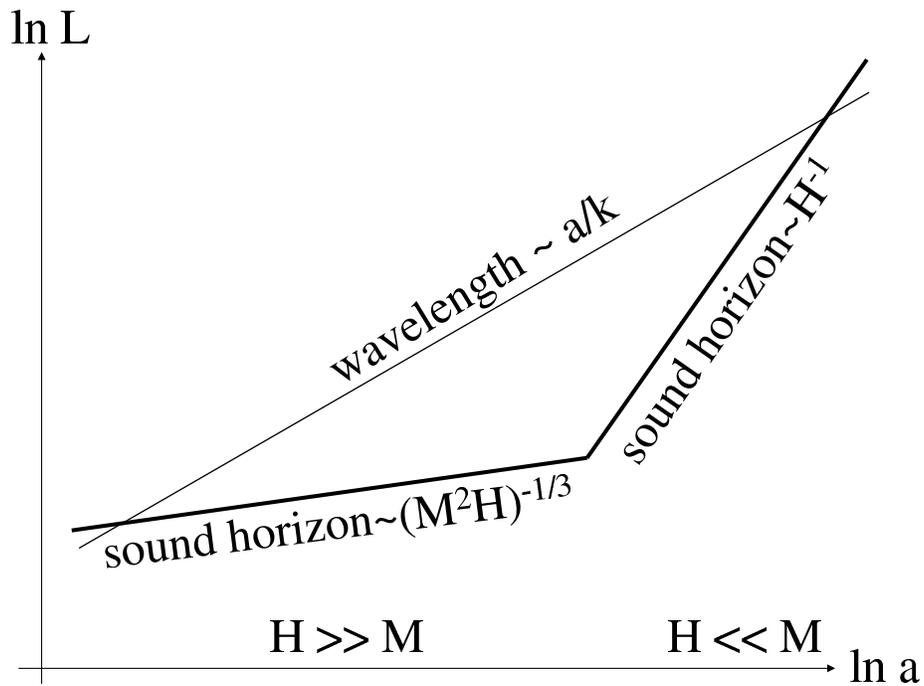}
\end{center}
 \caption{Physical wavelength ($\sim a/k$) exits the sound horizon 
($\sim (M^2H)^{-1/3}$) in the UV epoch ($H\gg M$) and re-enters the
horizon ($\sim H^{-1}$) in the IR epoch ($H\ll M$). In this figure, a
 power-law expansion $a\propto t^p$ with $1/3<p<1$ is supposed.}
\label{fig:scales}
\end{figure}

All necessary ingredients of the curvaton mechanism are already included
in the setup in the previous section. The value of $\Phi$ averaged over
a comoving size corresponding to the present horizon is expected to be
non-zero, (\ref{eqn:averagePhi}). Scale-invariant, super-horizon
fluctuations around the average are also generated in the UV epoch as
explained in the previous section. Once physical wavelengths of 
fluctuations exit the sound horizon $\sim (M^2H)^{-1/3}$ in the UV
epoch, both the average and the fluctuations of $\Phi$ are described by
the usual Klein-Gordon equation with mass $m$,
%
\begin{equation}
 \ddot{\Phi} + 3H\dot{\Phi}
  -\frac{\vec{\nabla}^2}{a^2}\Phi+ m^2\Phi = 0.
\end{equation}
The spatial gradient term becomes important only at and after the
horizon re-entry in the IR epoch~\footnote{Note that physical wavelength 
can re-enter the sound horizon only in the IR epoch, provided that the 
integral (\ref{eqn:integral}) converges. See Fig.~\ref{fig:scales}. 
Thus, once physical wavelength for the modes of interest exits the sound 
horizon in the UV epoch, the spatial gradient can be ignored all the way
down to the IR epoch, where the higher spatial derivative terms are
negligible.}. Therefore, the field $\Phi$ starts rolling and oscillating
around the origin when the Hubble expansion rate $H$ becomes as low as
$m$. Depending on the ratio $m/M$, this occurs in the UV epoch 
($m/M\gg 1$), the IR epoch ($m/M\ll 1$) or the intermediate epoch  
($m/M\sim O(1)$). Eventually, $\Phi$ decays into radiation and the 
fluctuations of $\Phi$ are converted to fluctuations of radiation and, 
hence, curvature perturbations.

Alternatively or supplementarily, we can invoke modulated decay of 
heavy particles or/and oscillating fields. Let us consider heavy
particles or/and oscillating fields after the scale-invariant
perturbations of $\Phi$ at scales of interest exit from the sound
horizon. Suppose that, at some point, energy density of the heavy
particles or/and the oscillating fields amounts to some fraction of  the
total energy density of the universe~\footnote{Energy density of the
heavy particles or/and the oscillating fields does not have to dominate
the universe.}. They must eventually decay before nucleo synthesis. If
the decay rate depends on the value of $\Phi$ (as in the case where
$\Phi$ is a modulus field) then fluctuations of $\Phi$ are converted to
fluctuations in decay products (or, equivalently, fluctuations in the
number of e-foldings) and, thus, to curvature perturbations.

In both scenarios, in the IR epoch, the $k^2/a^2$ term dominates the
dispersion relation and the corresponding sound horizon agrees with the
Hubble horizon. Therefore, in the IR epoch, curvature perturbations
re-enter the horizon as usual. See Fig.~\ref{fig:scales} for a schematic
picture of the exit from the sound horizon in the UV epoch ($H\gg M$)
and the re-entry to the horizon in the IR epoch ($H\ll M$). 
%

\section{Summary and discussions}
\label{sec:summary}

Based on Ho\v{r}ava-Lifshitz gravity, we have presented a simple
scenario for generation of almost scale-invariant, super-horizon 
curvature perturbations. The anisotropic scaling with dynamical
critical exponent $z=3$ implies that the amplitude of quantum
fluctuations generated in the early epoch of the expanding universe is
insensitive to the Hubble expansion rate and, thus,
scale-invariant. In Sec.~\ref{sec:scalar} we have analyzed mode
functions of a free scalar field to confirm the heuristic scaling
arguments in Sec.~\ref{sec:introduction}.

After leaving the sound horizon (corresponding to the $k^6/a^6$ term) in
the UV epoch, scale-invariant fluctuations of a scalar field are frozen
and then converted to curvature perturbations by the curvaton mechanism
or/and the modulated decay of heavy particles/oscillating fields. This
conversion can occur in either the UV epoch or the IR epoch (or the
intermediate epoch) without spoiling the scale-invariance of
fluctuations, as far as wavelengths of modes of interest are outside the
sound horizon. Note that the sound horizon in the UV epoch is far
outside the Hubble horizon. On the other hand, in the IR epoch, where 
$k^2/a^2$ term dominates, the corresponding sound horizon agrees with
the Hubble horizon and curvature perturbations re-enter the horizon as
usual. (See Fig.~\ref{fig:scales}.)

This scenario works as far as the integral (\ref{eqn:integral})
converges. For example, a power law expansion $a\propto t^p$ with
$p>1/3$ suffices and, thus, inflation is not required. Also, this
scenario does not rely on any additional assumptions such as the
detailed balance condition.

 So far, we have considered scalar-type cosmological
 perturbations. The analysis of tensor perturbations is essentially the
 same as that presented in Sec.~\ref{sec:scalar} for
 $\delta\Phi$. Therefore, the amplitude of tensor perturbations
 generated in the early epoch of the expanding universe is also
 insensitive to the Hubble expansion rate and, thus, scale-invariant.

 As discussed at the end of Sec.~\ref{sec:scalar}, for quantum
 fluctuations generated in the intermediate epoch, the power spectrum
 modestly depends on the background FRW evolution and is slightly
 tilted. This of course applies to both scalar- and tensor-type
 perturbations. 

\section*{Acknowledgements}

The author would like to thank Robert Brandenberger, Gianluca Calcagni,
Andrei Frolov, Roy Maartens and Takahiro Tanaka for useful comments. A
part of this work was done during YITP workshop ``Non-linear
cosmological perturbations'' (YITP-W-09-01). The work of the author was
supported in part by MEXT through a Grant-in-Aid for Young Scientists
(B) No.~17740134, by JSPS through a Grant-in-Aid for Creative Scientific
Research No.~19GS0219, and by the Mitsubishi Foundation. This work was
supported by World Premier International Research Center Initiative (WPI
Initiative), MEXT, Japan.

\section*{Note Added}

Refs.~\cite{Takahashi:2009wc,Calcagni:2009ar,Kiritsis:2009sh} also 
discuss cosmological perturbations. Here, we comment on those works. The
first version (v1) of ref.~\cite{Kiritsis:2009sh} does not include
discussions about scale-invariant cosmological perturbations. Actually,
the matter action presented in v1 does not include the $k^6$ term and
thus cannot lead to a scale-invariant spectrum from non-inflationary
epoch. After the present paper had appeared, the authors of
ref.~\cite{Kiritsis:2009sh} modified the matter action and added
discussions about scale-invariant cosmological perturbations from  
non-inflationary epoch in the second version (v2). Their conclusion in v2
is essentially the same as that in the present
paper. Ref.~\cite{Calcagni:2009ar} does not consider (sound) horizon
exit in the UV epoch and thus does not lead to scale-invariant spectrum
from non-inflationary epoch. Ref.~\cite{Takahashi:2009wc} investigates
tensor perturbations from a de Sitter phase ($a=-1/H\eta$) but does not
consider spectrum from non-inflationary epoch.


\end{document}